\documentclass[aps,twocolumn,prb,showpacs,10pt,floatfix,groupedaddress]{revtex4-1}
\usepackage{amssymb}
\usepackage{amsmath}
\usepackage{graphicx}
\usepackage{dcolumn}
\usepackage{bm}
\usepackage{textcomp}
\usepackage{color}

\begin{document}

\title{Mapping of strained graphene into one-dimensional Hamiltonians:
quasicrystals and modulated crystals}

\author{Gerardo G. Naumis}  
\email{naumis@fisica.unam.mx}
\author{Pedro Roman-Taboada}

\affiliation{Departamento de F\'{i}sica-Qu\'{i}mica, Instituto de
F\'{i}sica, Universidad Nacional Aut\'{o}noma de M\'{e}xico (UNAM),
Apartado Postal 20-364, 01000 M\'{e}xico, Distrito Federal,
M\'{e}xico}


\begin{abstract}

The electronic properties of graphene under any arbitrary uniaxial strain field 
are obtained by an exact mapping of the corresponding tight-binding Hamiltonian into an effective
one-dimensional modulated chain. For a periodic modulation, the system displays a rich behavior, 
including quasicrystals and modulated crystals with a complex spectrum, including gaps and peaks at the 
Fermi energy and localization transitions.  All these features are explained
by the incommensurate  or commensurate nature of the potential, which leads to a 
dense filling by diffraction spots of the reciprocal space in the former case. 
The essential features of strain are made specially clear by analyzing a
special momenta that uncouples the model into dimers. 

\end{abstract}


\maketitle
Graphene  is a two-dimensional (2D) carbon crystal \cite{Novoselov04}. This atom-thin elastic 
membrane  has amazing physical properties \cite{Novoselov04,Geim09,Novoselov11,Neto09}. Notably, 
graphene has the highest known interval of elastic response 
(up to $20$\% of the lattice parameter\cite{Lee08}). 
The tailoring of its electronic properties by 
controlled mechanic deformation is a field known as "straintronics" \cite{Pereira09a,Pereira09b,Guinea12,Zhan}. 
Also, graphene seems to be the ideal candidate to replace Si in transistors. 
But when graphene grows in top of a substrate with different 
lattice parameters or structure, strain and corrugation appear \cite{Vinogradov12}. 
The understanding of how strain affects the graphene's electronic properties is clearly a fundamental issue \cite{Kitt,FJ,KittE,Salvador,SalvadorSSC,Oliva13,Volovik} due to the complex 
self-similar structure of the reciprocal space. Such effect should be generated 
by growing graphene in top of a crystal with a slightly different lattice parameter, as is now
technically feasible\cite{Ponomarenko}. As is known, this
leads to a periodic strain \cite{Vinogradov12}. Then a quasiperiodic behavior should be obtained when the 
ratio of lattice parameters becomes incommensurate. Other two-dimensional materials  like
$MoS_2$ or $NiSe_2$ are expected to present the same effect \cite{Mos,Nise,Retana}.  


Let us start with a zig-zag graphene nanoribbon, as shown in Fig. 1, with a uniaxial strain applied in the $y$ direction.
Although our methodology can be applied for uniaxial strain in the zig-zag 
or arm chair directions, here we will concentrate only in one kind, since we want to bring out the essential
features of the model.

The new positions of the carbon atoms are  
$\bm{r}'=\bm{r}+\bm {u}(y)$, where $\bm{r}=(x,y)$ are the unstrained coordinates of the atoms and $\bm {u}(y)=(0,u(y))$ 
is the corresponding displacement. The electronic properties of graphene are 
well described by a one orbital next-nearest neighbor tight-binding Hamiltonian in a 
honeycomb lattice,  given by \cite{Neto09},   
\begin{equation}
H=-\sum_{\bm{r}'\!,n} t_{\bm{r}'\!,n} c_{\bm{r}'}^{\dag}
c_{\bm{r}' + \bm{\delta}_{n}'} + \text{H.c.}, \label{TBH}
\end{equation}
where $\bm{r}'$ runs over all sites of the deformed lattice
and $\bm{\delta}_{n}'$ are the corresponding vectors that point
to the three next nearest neighbors of $\bm{r}'$.
For unstrained graphene, $\bm{\delta}_{n}'=\bm{\delta}_{n}$ where,
\begin{equation}
\bm{\delta}_{1}=\frac{a}{2}(\sqrt{3},1), \ \
\bm{\delta}_{2}=\frac{a}{2}(-\sqrt{3},1),\ \
\bm{\delta}_{3}=a(0,-1).
\end{equation}

The operators $c_{\bm{r}'}^{\dag}$ and $c_{\bm{r}' + \bm{\delta}_{n}'}$
correspond to creating and annihilating electrons on lattice sites.  
The hopping integral $t_{\bm{r}'\!,n}$ depends upon strain, which induces
bond length changes that increase or decrease the overlap between wave functions. Such
variation with the distance can be calculated from\cite{Ribeiro09,Castro09} 
$t_{\bm{r}'\!,n}=t_0 exp[-\beta |\bm{\delta}_{n}'|/a -1)]$, where 
$\beta \approx 3$, 
$t_0 \approx 2.7 eV$ corresponds to non-strained pristine graphene, and 
$a$ is the bond length, which will be taken as $a=1$ in what follows.

\begin{figure}[h,t]
\includegraphics[scale=0.36]{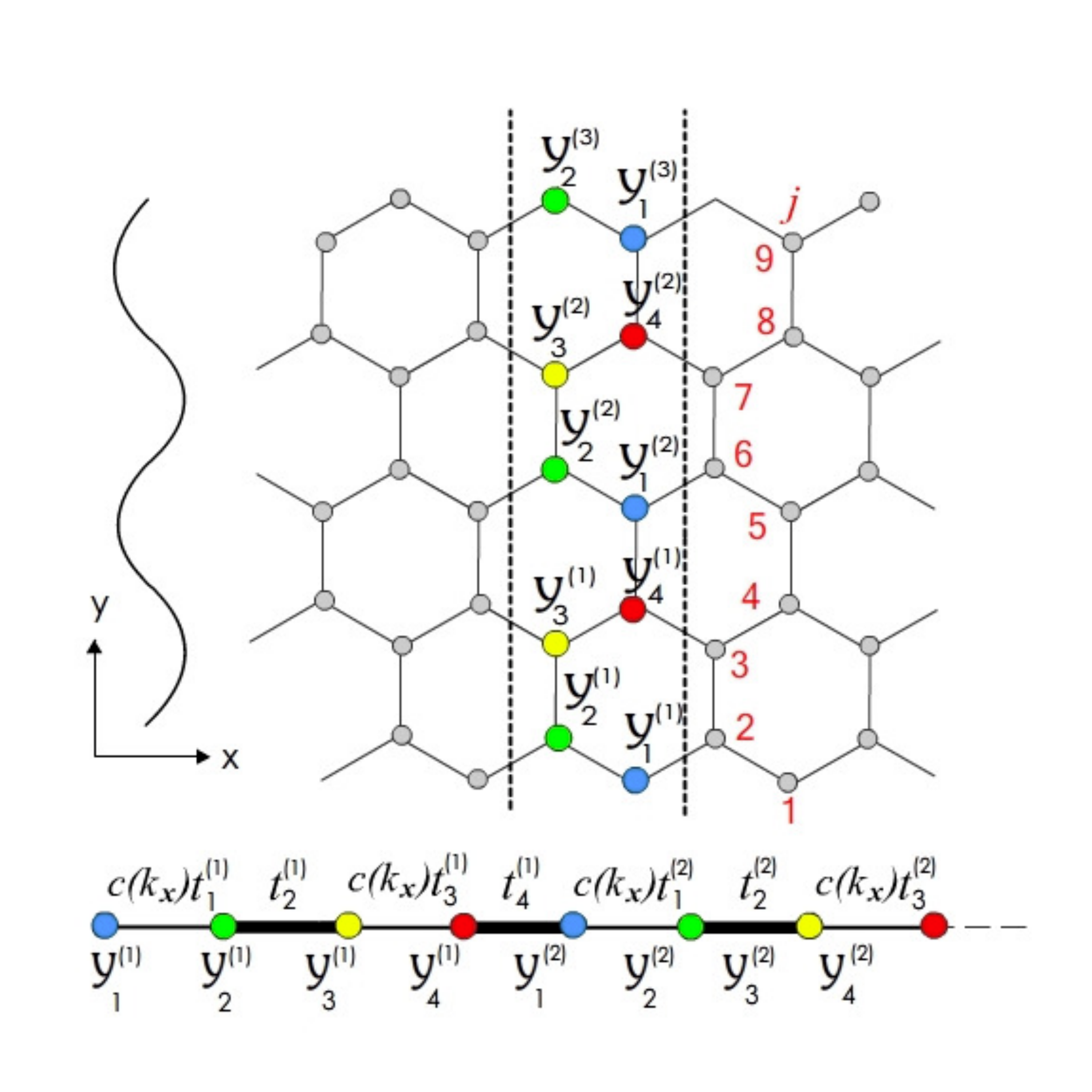}
\caption{\label{Map} (Color online) Mapping of zigzag strained graphene into a chain. The 
directions $x$ and $y$ are defined at the left. The strain in the $y$ direction is sketched 
out using a wavy curve, while the boundaries of the unitary cell in the $x$ direction are 
shown with dots. Inside the cell, four kinds of inequivalent sites appear (shown in different colors), 
denoted by $y_s^{(m)}$. The effective Hamiltonian of the zigzag path in the $y$ direction that joins 
sites $y_s^{(m)}$ can be mapped into the chain that appears below, where the label $j$ corresponds to the 
site along the zig-zag path as indicated. For the special momenta $k_x=\pi/\sqrt 3$, the model
breaks down into dimers, represented by bold links in the chain}
\end{figure}

For strain in one direction, we can map exactly the Hamiltonian  
into an effective one dimensional system as the  nanoribbon is made from cells of four non-equivalent 
atoms \cite{Roche} 
with coordinates $\bm{r}'=(x,y_{s}'^{(m)})$, where $s=1,2,3,4$ and $m$ denotes the number of the cell, as 
sketched out in Fig. 1. For graphene without strain, the positions in the $y$ direction are given by 
$y_1^{(m)}=3m$, $y_2^{(m)}=3m+1/2$, $y_3^{(m)}=3m+3/2$, and $y_4^{(m)}=3m+2$. On each of these sites, a strain field $u(y)$ is applied, 
resulting in new positions $y_{s}'^{(m)}=y_{s}^{(m)}+u_s^{(m)}$ where $u_s^{(m)}$ is a short hand 
notation for $u(y_s^{(m)})$. For uniaxial strain, the symmetry along the non-strained $x$ direction is not broken. 
Thus,  the solution of the 
Schr\"odinger equation $H\Psi(\bm{r}')=E\Psi(\bm{r}')$ for the energy $E$ 
has the form $\Psi(\bm{r}')=exp(ik_x x)\psi_{s}(m)$, where $k_x$ is the 
wave vector in $x$ direction and $\psi_s(m)$ is only a function of $y_{s}'^{(m)}$, where $s$ and $m$ label 
the sites along the zig-zag path in the vertical direction, as indicted in Fig. 1. Taking into account 
that for each bond that cross the dotted lines in Fig. 1, we need to add a phase $exp(\pm ik_x \sqrt 3 /2)$ for  
the wave-function, it is easy to obtain the following Schr\"odinger equation,
\begin{equation}
\begin{split}
E\psi_1(m)&=c(k_x)t_{1}^{(m)}\psi_{2}(m)+t_{4}^{(m-1)}\psi_{4}(m-1),\\
E\psi_2(m)&=t_{2}^{(m)}\psi_{3}(m)+c(k_x)t_{1}^{(m)}\psi_{1}(m),\\
E\psi_3(m)&=c(k_x)t_{3}^{(m)}\psi_{4}(m)+t_{2}^{(m)}\psi_{2}(m),\\
E\psi_4(m)&=t_{4}^{(m)}\psi_{1}(m+1)+c(k_x)t_{3}^{(m)}\psi_{3}(m),\\
\end{split}
\end{equation}
where $c(k_x)=2\cos(\sqrt 3 k_x/2)$ and
$t_{s}^{(m)}=t_0 exp[-\beta (u_{s+1}^{(m)}-u_s^{(m)})\delta_{s+1,s}^y)]$.
Here $\delta_{s+1,s}^y$ denotes the $y$ components of each of the three vectors
$\bm{\delta}_{1},\bm{\delta}_{2},\bm{\delta}_{3}$ that join sites with $y$ coordinates $y_s^{(m)}$ and $y_{s+1}^{(m)}$ 
for  unstrained graphene. In this formula, one needs to apply the conditions $y_5^{(m)}=y_1^{(m+1)}$ and 
$y_0^{(m)}=y_4^{(m-1)}$ at 
the boundary of each cell. Furthermore, the sequence of $y_s^{(m)}$ can be written as 
$y(j)=[3j+(1-(-1)^{j})/2)]/4$ where $j$ is an integer that labels the site number along the zigzag path 
in the $y$ axis, given by $j=4(m-1)+s$. Finally, one can write a Hamiltonian $H(k_x)$ without 
any reference to cells of four sites, 

\begin{equation}
H(k_x)=\sum_{j} \left[ t_{2j} c_{2j+1}^{\dag} c_{2j}+c(k_x)t_{2j+1} c_{2j+2}^{\dag} c_{2j+1} \right]
\label{Hkx}
\end{equation}
with $t_{[4(m-1)+s]}=t_{s}^{(m) }$. This gives

\begin{equation}
t_{j}=t_0 exp\left[-\beta \frac{3+(-1)^{j+1}}{4}(u_{j+1}-u_{j})\right].
\label{tsm}
\end{equation}
where it is understood that $u_j$ is just the displacement of the $j$-th atom along the vertical 
zig-zag path, i.e., $u_j\equiv u_s^{(m)}$.   Now $H(k_x)$ describes a chain for 
any arbitrary uniaxial strain, as indicated in Fig. 1.

The exact mapping can serve as a test for approximate theories of strain in graphene. Consider for example an 
oscillating strain 
$u(y)=(2/3)(\lambda /\beta)\cos[(8 \pi/3)\sigma (y-1/2)  +\phi]$, of the type expected when graphene grows 
on top of a material with a different lattice \cite{Vinogradov12}.    

\begin{center}
\begin{figure}[!ht]
\includegraphics[width=8.6cm,height=8.6cm]{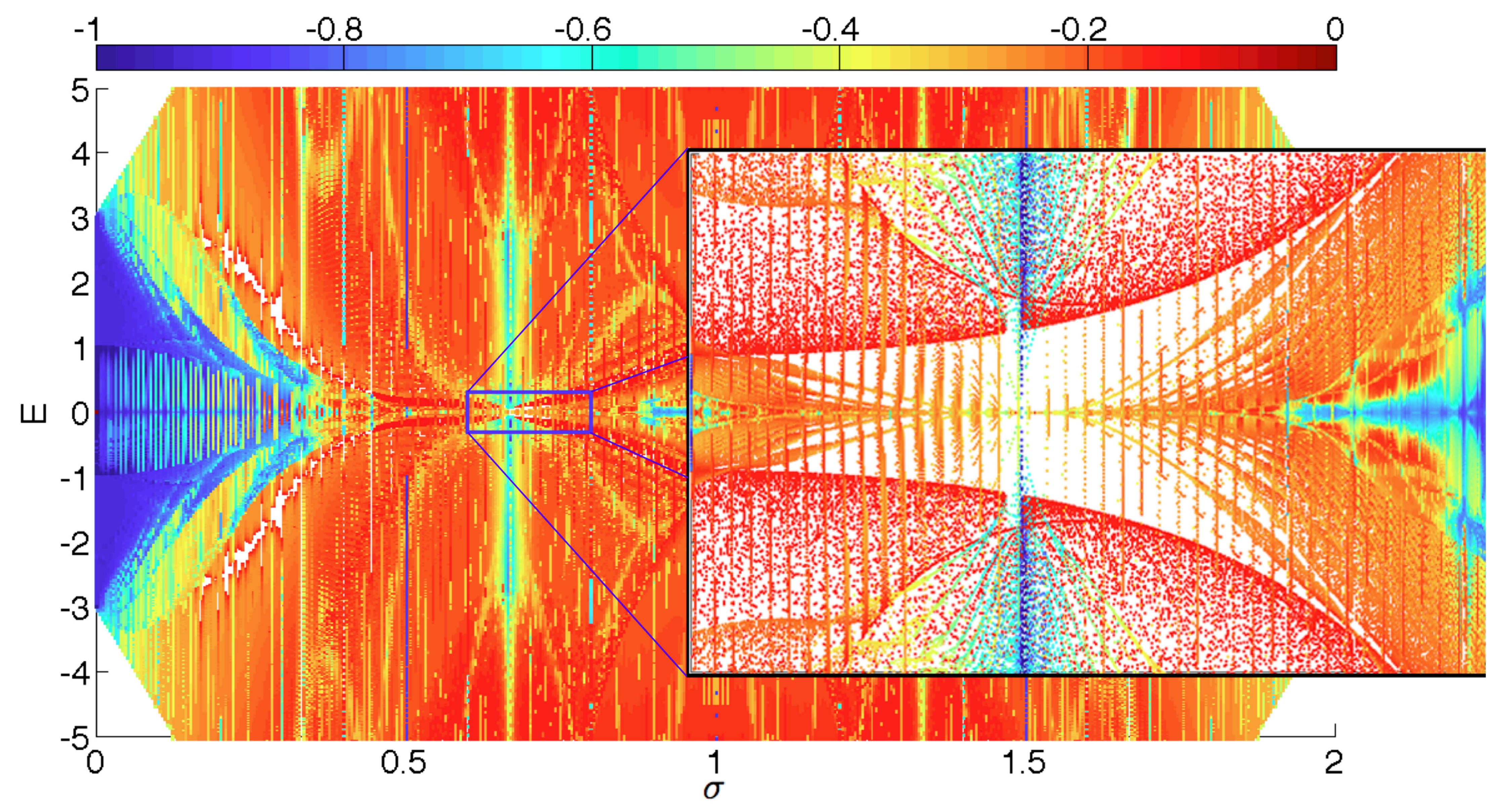}
\caption{\label{Spectrum} (Color online) Spectrum as a function of $\sigma$ 
for $\lambda=2$ and $\phi=(4/3)\pi \sigma$ considering the exponential 
dependence of $t_{\bm{r}'\!,n}$, obtained by solving the Schr\"{o}dinger's equation 
for a system of $200$ atoms, using $300$ grid points for sampling  $k_x$ and with periodic 
boundary conditions. The inset presents a blow up near zero energy. The 
different colors represent the localization participation ratio $\alpha(E)$.}
\end{figure}
\end{center}

Figure \ref{Spectrum} shows the complex spectrum of $H$ as a function of $\sigma$, revealing a 
behavior that is akin to the Hofstadter butterfly that appears in the Harper model \cite{Hofstadter76}. The 
most surprising result is the appearance of gaps around the Femi level $E=0$ for some values of $\sigma$. We can get
a better understanding by using a linear approximation for $t_{\bm{r}'\!,n}$, assuming a small strain as 
usual in straintronics. Under such approximation, Eq. (\ref{tsm}) becomes
\begin{equation}
\frac{t_{j}}{t_0}=1+\lambda \xi(j+1) \sin(\pi \sigma \xi(j)) \sin(2 \pi \sigma j+\phi) 
\label{tvalue}
\end{equation}
where $\xi(j)=1+\left[(-1)^{j}/3\right]$.

\begin{figure}
\includegraphics[width=8.5cm,height=7.4cm]{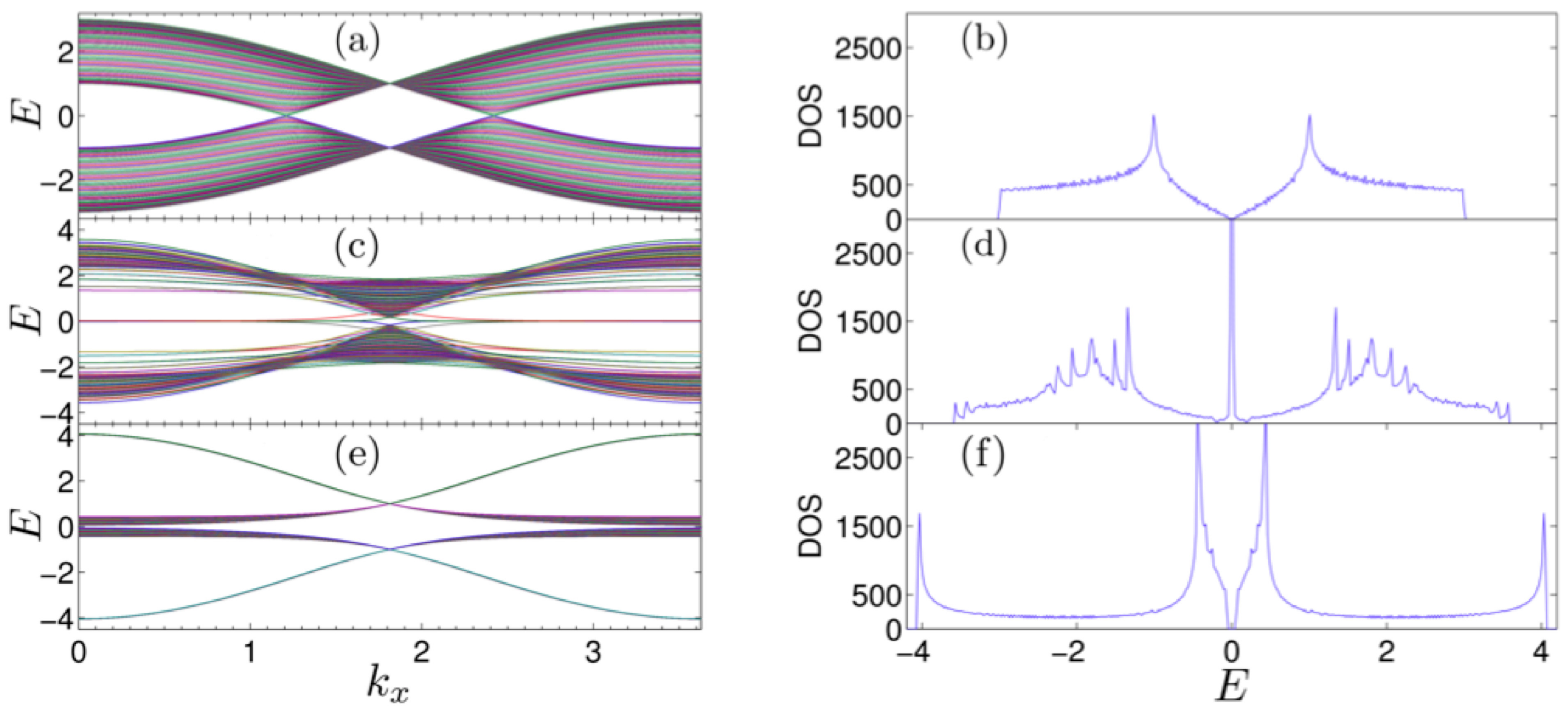}
\caption{\label{DOS} Band structure (left column) and density of states (right
column) using $\phi=(4/3)\pi \sigma$ and $\lambda=2$ for, (a) and (b) unstrained graphene lattice, 
(c) and (d) strained graphene with $\sigma \approx 3\tau/4$,
(e) and (f) strained graphene with $\sigma=3/4 $. Spikes appear for cases (c) and (d), while a
gap is seen in (e) and (f) at $E=0$. Observe in (e) and (f) how 
the DOS is similar to linear chains perturbed by a small interaction.}
\end{figure}

\begin{center}
\begin{figure}
\includegraphics[width=8.6cm,height=6.3cm]{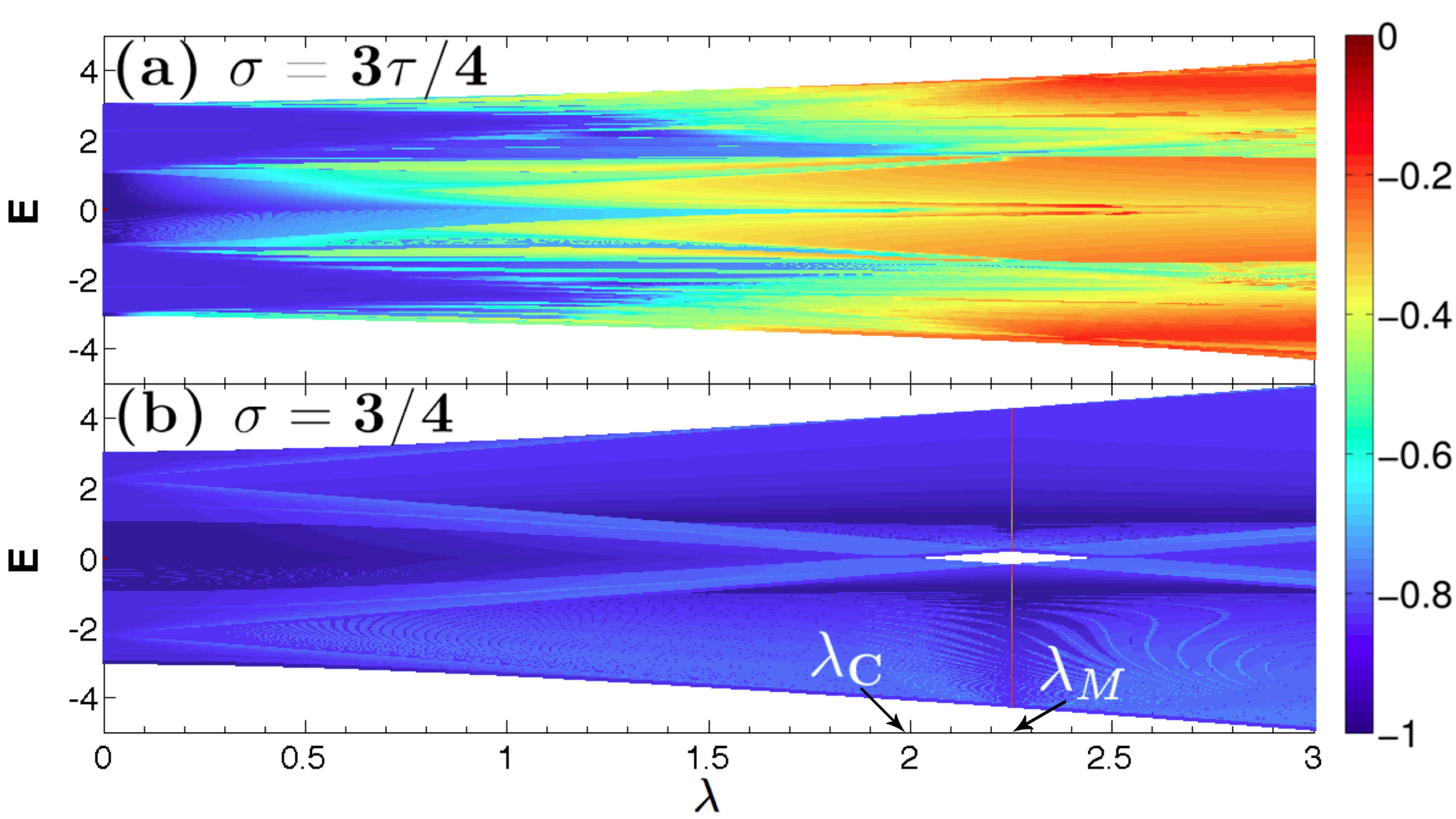}
\caption{\label{Gapopening} Energy spectrum of graphene under uniaxial sinusoidal strain in the
linear approximation as a function of $\lambda$ for (a) $\sigma=(4/3)\pi \tau$ and (b) $\sigma=3/4$. 
In (b), a gap opens for $\lambda>\lambda_{C}=9\sqrt(3)/8$, while for $\lambda=9/4$ the system 
breaks down into disconnected strips of triangular cells. The colors represent the localization 
participation ratio $\alpha(E)$. Notice the transition at $\lambda=\lambda_{M}$. The phase
was taken as $\phi=(4/3)\pi \sigma$, and periodic boundary conditions were used.}
\end{figure}
\end{center}

The resulting Hamiltonian describes one dimensional quasicrystals for irrational $\sigma$ , and  modulated 
crystals for  rational $\sigma$ . Although the model resembles an off diagonal Harper model\cite{Harper55}, there 
is an important extra modulation provided by $\xi(j+1)\sin(\pi \sigma \xi(j))$. In Fig. \ref{DOS} we present 
the resulting bands as a function of $k_x$ and the corresponding density of states (DOS) for  
$\sigma=0$ (pure graphene), $\sigma=3\tau/4$ and $ \sigma=3/4$, where $\tau$ is the 
golden ratio $\tau=(\sqrt 5+1)/2$. Several interesting features are observed. The first 
is the disappearance of the Dirac cone for cases (c) and (e), observed around $E=0$ for 
pure graphene. In case (c), degenerate states appear at $E=0$ and the DOS is spiky. On the 
other hand, in case (f) the DOS is smooth. Only the Van Hove singularities observed for $E=\pm 1$ in 
pure graphene move and split in two. 
It is also interesting the behavior of the spectrum as a function of $\lambda$ for a given $\sigma$. In
Fig. \ref{Gapopening} (a) and (b), we present the cases  $\sigma=3\tau/4$ and $\sigma=3/4$. For $\sigma=3/4$, a gap
opens above a certain critical $\lambda_{C}$, while for $\sigma=3\tau/4$, no gaps are seen. Let us  
explain this rich  behavior.

An analysis of Fig. \ref{DOS} (e) suggests that for $\sigma=3/4$, the behavior is akin to a system of two 
disconnected chains. Such analysis is confirmed by evaluating Eq. (\ref{tsm}) using $\sigma=3/4$. In 
this particular case, the strain has the same period as the four site cells, thus $t_{s}^{(m)}$ turns out to be 
independent of $m$, and $t_{s}^{(m)}=1-4(\lambda/3) \sin (3 \pi s/2)$. The corresponding band edges are given
by a matrix of $4\times4$ whose solutions, in terms of the parameters $\lambda$ and $k_x$, are  
$E(\lambda,k_x)=\pm \Big( \sqrt{1+(8\lambda/9)^{2}\cos^2{(\sqrt3 k_x/2)} }  \pm 2\cos{(\sqrt3 k_x/2)} \Big)$. 
A gap opens when $\lambda>\lambda_{C}=9\sqrt3/8$. For $\lambda=\lambda_{M}=9/4$, 
the system behaves as two disconnected strips of triangular cells, explaining the observed spectrum of Fig. \ref{DOS} 
(e). The gap ($\Delta$) goes as $\Delta \propto (\lambda-\lambda_{C})$ as confirmed by Fig. \ref{Gapopening} (b). 

Figures \ref{DOS} (c) and \ref{Gapopening} (a) are even more interesting. Here the strain is 
incommensurate with the four site cell period. The system is thus quasiperiodic. As is well known, perturbation theory 
can not be used at any order, since the  problem is akin to the small divisor problem  due to the dense appearance of 
diffraction peaks in reciprocal space\cite{Steinhardt}. This fact is important since if a Fourier expansion of 
the operator $c_{\bm{r}'}$ is  performed 
as $c_{\bm{r}'}=\sum_{\bm{k}}exp[\bm{k}\cdot(\bm{r}+\bm{u}(\bm{r}))]c_{\bm{k}}$, where $\bm{k}$ is 
a reciprocal vector, then  one needs to  consider a dense distribution\cite{Steinhardt,Naumis08} in 
$\sum_{\bm{k}}exp[\bm{k}\cdot\bm{u}(\bm{r}))]$ for incommensurate cases. This explains the spiky DOS, 
since for each diffraction spot, a singularity appears \cite{Kittel,Naumis08}. To overline this, let us 
work out a particular example.

For the value $k_x=\pi/\sqrt 3$, we have that $c(k_x)=0$. This is valid for any $\lambda$ or $\sigma$. 
The corresponding Hamiltonian $H(k_x=\pi /\sqrt3)$  given by Eq. (\ref{Hkx}) becomes just a model for disconnected dimers,
represented in the chain of Fig. \ref{Map} as bold lines.
The eigenvalues are obtained from an effective $2\times2$ matrix, from where $E(k_x=\pi/\sqrt 3)=\pm t_{2l}$,  
with $l$ an integer. Using Eq. (\ref{tvalue}), the eigenvalues are  
$E(k_x=\pi/\sqrt 3)=\pm [1+(2/3)\lambda \sin (4\pi \sigma/3) \sin (4 \pi \sigma l+\phi)]$.
In the case of unstrained graphene, $E(k_x=\pi/\sqrt 3)=\pm 1$. These two values correspond to the highly degenerate peaks
observed in the DOS of Fig. \ref{DOS} (b). Each peak has a degeneracy $N/2$, where $N$ is  
the number of atoms in the zig-zag path. These peaks are associated with a Van Hove singularity, since
standing waves due to diffraction appear\cite{Kittel,Naumis08,BarriosVargas201323}. 
For $\sigma=3/4$, $E(k_x=\pi/\sqrt 3)=\pm 1$. The 
degeneracy remains, as seen in  \ref{DOS} (e), although it does not produce peaks because all other states are
also highly degenerate. However, for 
irrational $\sigma$, the factor $ \sin (4 \pi \sigma l+\phi)$ behaves as a pseudorandom number generator which fills
in a dense way the interval \cite{Steinhardt} $[-1,1]$. The degeneracy is thus lifted. The spectral type is pure point 
and contained in the intervals $[-1-2\lambda/3,-1+2\lambda/3]$ and $[1-2\lambda/3,1+2\lambda/3]$, leading to a gap
of size $4\lambda/3$ if $\lambda<3/2$. The splitting is evident at the middle of $k_{x}$ axis in Fig. \ref{DOS} (c), and
when compared with Fig. \ref{DOS} (a) and (e). What happens to the wave function's localization? For irrational $\sigma$, the
eigenfunctions are localized in dimers on the $y$ direction. Obviously, since all $E(k_x=\pi/\sqrt 3)$ are different,
an infinite number of reciprocal vectors are needed to generate the corresponding wave functions. Thus, 
even in this simple case the usual perturbation theory breaks down. 
However, for rational $\sigma$, the eigenvalues are degenerate. Any linear combination of the wave function
in dimers is a solution, leading to delocalized states around $k_x=\pi/\sqrt 3$. Such behavior is revealed by calculating
the normalized participation ratio, defined as \cite{Naumis07},
\begin{equation}
\alpha(E)=\frac{\log \sum _{j=1} ^{N}|\psi(j)|^{4}}{\log N}.
\end{equation}

The factor $\alpha(E)$ is
a measure of localization. In Fig. \ref{Spectrum} and \ref{Gapopening}, the colors indicate the value of
$\alpha(E)$. For Fig. \ref{Spectrum}, a fractal behavior reveals how localization depends on the number theory
properties of $\sigma$. In Fig. \ref{Gapopening} (b) the case $\sigma=3/4$ does not present appreciable 
changes, as expected from the previous discussion. Only at  $\lambda=\lambda_{M}$ there is a 
localization transition as a
consequence of the breaking into disconnected chains, 
leading to the vertical red line observed in Fig. \ref{Gapopening} (b). The case $\sigma=3\tau/4$ shows the expected localization
around $E=\pm 1$ as $\lambda \rightarrow \infty$.

Finally, it is worthwhile mentioning how some of the observed effects  are related with the 
zigzag states reported in graphene nanoribbons\cite{Nakada, Otrosnakdada1, Otrosnakdada2} and 
topological states\cite{Satija13}.  
In particular, the DOS in the irrational case resembles the case of narrow graphene nanoribbons\cite{Nakada}.
The reason for this is simple. For irrational $\sigma$, there are sites $j$ in which $t_j \approx 0$, since
$t_j$ mimics a random number generator. In such sites, the lattice is almost decoupled in the $y$ direction,
producing many effective nanoribbons of different widths. This leads to singularities that are strikingly similar
to narrow nanoribbons, as observed in \ref{DOS} (d). In fact, a similar phenomena happens for rational $\sigma$
and big $\lambda$. For example, if $\sigma=3/4$ and $\lambda=\lambda_D$, 
$t_j$ is zero at the end of the unitary one dimensional cell and we obtain many effective nanoribbons, but
this time all with the same four atom width. In a similar way, the states at the Fermi energy can be explained
in many different way: as zigzag states \cite{Nakada} due to an effective decoupling
in nanoribbons, as an imbalance in the number 
of atoms in each bipartite lattice\cite{BarriosVargas201323} or as strictly confined states\cite{BarriosVargas201323}.
These states have a toplogical nature, as we have verified by changing $\phi$ and using different 
boundary conditions.

In conclusion, we have provided an exact mapping into a one dimensional chain for any uniaxial 
strain in graphene. For a periodic strain, effective quasiperiodic or modulated crystals systems 
were obtained. Due 
to the dense nature of  the reciprocal space, the spectrum and localization properties 
presented a fractal pattern. Gaps, singularities and localized states were observed. These 
features can not be predicted by simple perturbation theory techniques. The quasiperiodic 
nature of the problem found here, suggests  the paramount importance of disorder due to 
the intrinsic instability of such spectra \cite{Naumis96, Naumis98, Naumis05, lopez93} and the possibility of building
equivalent superlattices \cite{Nava}.  
In future work, we will study edge states, since they are expected to present a
nesting of topological length scales, given by the Chern numbers,
within a fractal pattern, as observed in the  Harper and Fibonacci models\cite{Satija13}.

This work was supported by DGAPA-PAPIIT IN-$102513$, and by DGTIC-NES center.



\end{document}